\documentstyle[12pt]{article}

\begin{document}

		\begin{titlepage}
\setcounter{page}{1}

\date{}
\title{\small\centerline{\hfill August 1993 }
\bigskip\bigskip
{\Large\bf Testing  the handedness of a heavy $W^{\prime}$
         at future hadron colliders}\bigskip}

\author{Mirjam Cveti\v c,  Paul Langacker and Jiang Liu\\
\normalsize
\em Department of Physics\\
\normalsize
\em University of Pennsylvania\\
\normalsize
\em Philadelphia, PA 19104\\
\normalsize
UPR-0579T}
\maketitle
\vfill

\begin{abstract}\noindent\normalsize
We show that the associated
production $pp\rightarrow W^{\prime }W$,
and the rare decay $pp\rightarrow  W'\rightarrow \bar\ell\ell W$
are  useful tests of $W'$ couplings to fermions
at future hadron colliders.
For $M_{W'}\sim (1 - 3)$ TeV they  would allow a
clean determination on whether
the $W'$ couples to $V-A$ or $V+A$ currents.
As an illustration a model
in which the $W^{\prime\pm}$ couples only
to $V-A$ currents is contrasted to the left-right symmetric
models which involve $V+A$ currents.
\end{abstract}

\vfill
\centerline{PACS \# 12.15, 12.10, 11.15\hfill }

\thispagestyle{empty}
\end{titlepage}


Many types of new physics, including some grand unified  and
superstring theories, predict the existence of additional
charged and neutral gauge bosons ($W', Z'$).  While their masses are
{\it a priori} arbitrary, it is at least possible that they
may be in the experimentally accessible range of a few TeV
\cite{rfi}.

The present direct and indirect limits on additional gauge
bosons are very model dependent.  The bounds on the  mass of  a
new $Z'$ are $160-400$ GeV
\cite{rfv,rfii,rfiii,delA},
although the limits are stronger, e.g.,
$500-1000$ GeV, in some models in which the $Z'$ mass and the $Z-Z'$
mixing are related.
In the version of
left-right symmetric models \cite{rfip}
with equal left- and right-handed gauge couplings
and magnitudes of quark-mixing matrix elements one has the stringent
limit $M_{W'}>1.4\hbox{~TeV} $ from
the $K_L-K_S$ mass difference \cite{rfiip}.
In general left-right models, however,  one has the
weaker limit \cite{rfivp}
$g_L M_{W'}/g_R>300\hbox{~GeV}$.
Stronger limits follow from CP violation unless
there is fine tuning \cite{rfLoWy}.

Heavy $Z'$ and $W'$ can be produced and detected by their leptonic
decays at the
Large Hadron Collider (LHC) and Superconducting Super Collider (SSC)
for masses up to $\sim 5$ TeV \cite{rfi},
\cite{rfx}-\cite{HRII}.
To identify the origin of such bosons, more detailed
diagnostic probes of their couplings will be needed.
Recent detailed studies \cite{rfxiiip}-\cite{25a}
have demonstrated
that the rare decay
process \cite{rfxiiip,rfxiv}
$Z'\to \bar f_1f_2 V\ (V=W,Z)$, where
$f_{1,2}$ are ordinary fermions, the associated production \cite{asso}
$pp\to Z'V $, and the  rapidity distribution in
$pp\to Z'\to \ell^+\ell^-$ \cite{25a}
 are useful diagnostics of the
$Z'$ couplings to the ordinary fermions.

Another clean probe for the gauge couplings of $Z'$ and $W'$ is the
forward-backward asymmetry \cite{LRR}.  For
$pp\to Z'\to \ell^+\ell^-\ (\ell=e$ or $\mu$) and
$pp\to W^{\prime\pm}\to \ell^{\pm}\nu_{\ell}$ the
asymmetries can distinguish between different models
for $M_{Z'(W')}$ up to a few TeV, and test some
combinations of the couplings of $Z'$ and $W'$ to quarks and leptons.
However, the forward-backward asymmetry for $W^{\prime\pm}$ does
not distinguish $V+A$ couplings from $V-A$.
Although the most likely extension of the standard model involving
a $W'$ is the left-right symmetric model \cite{rfip}
with $V+A$ couplings,   it is possible
to construct viable models with $V-A$ couplings as well \cite{GJS}.
It is therefore  important to be able to distinguish $V+A$ from $V-A$.

Possibilities for distinguishing the handedness of $W'$
have been recently pointed out \cite{rfxiv,rfix,asso}.
The basic idea is that the ordinary $W^{\pm}$ has only $V-A$ couplings, which
acts as a filter for testing the handedness of $W^{\prime\pm}$.
For example, if $W'$ has only $V+A$ couplings,
the decay
$W^{\prime\pm}\to W^{\pm}\ell^+\ell^-$ will not occur at the lowest order
except for small corrections from lepton masses.
For the same reason the process $pp\to W^{\prime\pm}W^{\mp}$
would be strongly suppressed if $W'$ has the opposite
handedness as $W$:  in the left-right-symmetric model the
suppression is proportional to the square of
the $W'-W$ mixing angle or to the ratio $m_f^2/M_W^2$, where
$m_f$ is a small fermion mass.

On the other hand, if the $W'$ couples to $V-A$ currents these
processes would not be suppressed by the mismatch of the
handedness.  In this paper we will examine this
possibility in more detail.  We will show that the
number of events in the $V-A$ case can be sufficiently large to
allow a clean determination of the handedness of
a $W'$ with mass of the order of
$(1-3)$ TeV.

As an illustration we consider a theory \cite{GJS} in
which $W'$ couples to $V-A$ currents.  This is an `un-unified' theory
of weak interaction
with a gauge structure $SU(2)_{q}\times SU(2)_{\ell}\times U(1)_Y$,  in which
the left-handed quarks and leptons
transform as doublets of their own $SU(2)$.  One set of linear combinations
of the gauge bosons of $SU(2)_q$ and $SU(2)_{\ell}$ give the standard
$W$ and $Z$, and the other become $W'$ and $Z'$.
In this model both $W'$ and $Z'$ couple to $V-A$ currents.
While this model was
originally proposed as an alternative
to the standard electroweak model
with relatively light $W'$ and $Z'$,
for our purpose we only consider situations in which
the extra gauge bosons are heavy, i.e., $M_{W'}, M_{Z'}\geq1$ TeV.
Then to leading order of $M_W^2/M_{W'}^2$ one finds $M_{W'}= M_{Z'}$.
Neglecting fermion mixings the charged current interaction is given by
\begin{eqnarray}
{\cal{L}}_{\rm CC}=-{g\over 2\sqrt 2}
    &\Bigl\{&
    \Bigl[W^-_{\mu} + \cot \phi W^{\prime -}_{\mu}\Bigr]\bar
u\gamma^{\mu}(1-\gamma_5)
     d\nonumber\\
&+& \Bigl[W^-_{\mu}-\tan\phi W^{\prime
-}_{\mu}\Bigr]\bar\ell\gamma^{\mu}(1-\gamma_5)
    \nu_{\ell}\Bigr\},\label{CC}
\end{eqnarray}
where $\tan\phi=g_{\ell}/g_{q}$  and $g_{\ell(q)}$ is the
gauge coupling constant of $SU(2)_{\ell(q)}$. To have a meaningful
perturbation calculation in what follows we consider  that $\phi$
is not close to $0$ or $\pi/2$.
The neutral current interaction of $Z$ has the standard form, whereas
that of $Z'$ is
\begin{eqnarray}
{\cal{L}}_{NC}(Z')=-{g\over 4}Z'_{\mu}&\Bigl\{&
       \cot\phi\Bigl[\bar u\gamma^{\mu}(1-\gamma_5)u - \bar d\gamma^{\mu}
           (1-\gamma_5)d\Bigr]\nonumber\\
      &-& \tan\phi\Bigl[\bar\nu_{\ell}\gamma^{\mu}(1-\gamma_5)\nu_{\ell}
        -\bar\ell\gamma^{\mu}(1-\gamma_5)\ell\Bigr]\Bigr\}.\label{NC}
\end{eqnarray}
The coupling constant for the trilinear
$W'Z'W$ vertex is $g$, those involving
a $W'ZW$ or $W'\gamma W$ vertex are further suppressed
by the ratio $M_W^2/M_{W'}^2$.  Gauge invariance relates
the couplings in (\ref{NC}) and the trilinear gauge interactions,
resulting in  a destructive interference
for the physical processes discussed below.

The heavy charged gauge boson $W'$, assuming its existence,  can be produced at
future
hadron colliders (SSC and LHC) and can be detected via the
resultant leptonic decays
$pp\to W'\to \bar{\ell}\nu_{\ell} (\bar\nu_{\ell}\ell)$.
For given $W'$ couplings the total cross section
$\sigma(pp\to W')$ can be computed quite accurately.
The cross sections are given in
\cite{LRR,Bar,DuLa,HRII}. For definiteness, we assume that the
neutrinos to which the $W'$ couples are massless or light.
This is the case for the un-unified model \cite{GJS},
for which $\nu_{\ell}$ is the ordinary neutrino, and in some
versions of the left-right symmetric model.  The same ideas
would apply to models involving heavy (e.g. Majorona) neutrinos.

We first address the associated production.
In the un-unified model there are two tree-level
graphs (Fig. 1).
Contributions
from these two graphs are equally important.
In fact,
gauge invariance requires that they
interfere destructively to enforce
unitarity.
The squared amplitude
for the quark process $q\bar q\to W'W$
averaged (summed) over initial (final) polarizations is
\begin{eqnarray}
{d\sigma_{W'W}\over dt} = {g^4\cot^2\phi\over 16\pi s^2}{\cal{M}},
   \label{sigma}
\end{eqnarray}
where
\begin{eqnarray}
{\cal{M}} =
    &-& {1\over 4t^2}\Bigl[ 3t^2 + t(s + M_{W'}^2 + M_W^2)
     + M_{W'}^2M_W^2\Bigr]\nonumber \\
   &+& {1\over (s-M_{W'}^2)}\Bigl[
       -{M_{W'}^2\over 2} + {M_W^2\over 8} + {M_W^4\over 8 M_{W'}^2}
        -{M_{W'}^4\over 2t} -{M_{W'}^2M_W^2\over t} -
        {tM_W^2\over 16M_{W'}^2} -t\Bigr]\label{M}\\
   &+& {1\over (s-M_{W'}^2)^2}
    \Bigl[-{M_{W'}^2\over 2} - {7M_{W'}^2M_W^2\over 8}
          + {M_W^4\over 16} + {t M_W^2 \over 2}
          + { t M_W^4 \over 16 M_{W'}^2}
          -{t^2\over 2}
          - { t^2M_W^2 \over 16 M_{W'}^2}\Bigr],\nonumber
\end{eqnarray}
and $s, t$ are the Mandelstam variables.

The total
cross section for $\sigma_{W'W}$ is obtained in a straightforward manner
using the quark distribution functions of Ref. \cite{EHLQ}.
We define the cross section for $pp\to W'W$ as the sum over
$W^{\prime +}W^-$ and $W^{\prime -}W^+$.
For a one year ($10^7\ s$) run at the LHC (SSC) with the projected
luminosity of  $10^{34}(10^{33})\ cm^{-2}\ s^{-1}$, the
number of events
along with typical statistical errors for
the process $pp\to W'W$, with $W'$ subsequently decaying
into $\bar\ell\nu_{\ell}$ and $\bar\nu_{\ell}\ell$
($\ell= e$ and $\mu$) is presented in Table 1.
In obtaining these results we have assumed for simplicity  that the $W'$
only decays to ordinary fermions,  with
the leading term of its total rate given by
\begin{eqnarray}
\Gamma(W'\to\bar ff')={g^2M_{W'}\over 16\pi}\Bigl[
    \tan^2\phi + 3\cot^2\phi\Bigr].\label{Gammatotal}
\end{eqnarray}
These numbers are presented only for illustration.  They
should be contrasted with number zero, which
would be the result if
the $W'$ coupled to $V+A$ currents.
On average the numbers for the $W'W$ associated production are
about two orders of magnitude larger than those from the
$Z'W$ associated production \cite{rfxiv}.  One major reason is that
in this model $W'$ has a larger coupling.  Thus, the
signal is still significant even for $M_{W'}= 3$ TeV.

The production of $W'W$, with $W'$ subsequently
decaying into leptons and $W$ into hadrons
are clean events without major background.
The standard-model background from $pp\to WW$ with one
$W$ decaying into leptons can be cleanly eliminated
at a loss of only a few percent of the signal by requiring
the transverse invariant mass of the lepton system to be larger than
$90$ GeV.

We now turn to the rare decay process $W'\to \bar\ell\ell W$.
There are two tree-level graphs displayed in Fig. 2, and
their contributions are equally important.
Averaging (summing) over
the initial (final) polarizations of the squared amplitude we find
\begin{eqnarray}
{d\Gamma(W'\to \bar\ell\ell W)\over d\omega} =
{g^4\tan^2\phi\over 24 (2\pi)^5 M_{W'}^3 t^2(s-M_{W'}^2)^2}
\delta^4(P_{W'}-P_{W}-P_{\ell}-P_{\bar\ell}){\cal{M'}},
\label{rate}
\end{eqnarray}
where  $d\omega=(d^3\vec{P}_{W}/2P_{W}^0)
                (d^3\vec{P}_{\bar\ell}/2P_{\bar\ell}^0)
                (d^3\vec{P}_{\ell}/2P_{\ell}^0)$,
$t=(P_{W'}-P_{\ell})^2, s=(P_{\bar\ell} + P_{\ell})^2$, with
$P_{W'}, P_{W}, P_{\bar\ell}$ and $P_{\ell}$ referring to the momenta of
the corresponding particles, and
\newpage
\begin{eqnarray}
{\cal{M}}'= &\ & 16 t^2M_{W'}^2\Bigl[M_{W'}^2(s - M_W^2)
                   - t(s - M_{W'}^2)\Bigr]\nonumber
\\
           &+& 4 t M_{W'}^2\Bigl[3 M_{W'}^4M_W^2 - 3t ( s + M_{W'}^4)
                                 - s(M_{W'}^4 +s M_W^2  + s^2)\Bigr]\nonumber\\
 &+& 4 M_{W'}^2\Bigl[M_{W'}^2(s^2 + M_{W'}^4) (t-M_W^2)
           - 2sM_{W'}^2M_W^2 (t-M_{W'}^2  ) - 2t^4\Bigr]\label{M'}\\
&+& t^2 M_W^2\Bigl[2s(M_{W'}^2 + M_W^2) - t(t+s-9M_{W'}^2)
           + M_W^2(t-M_{W'}^2)\Bigr].\nonumber
\end{eqnarray}
Due to the destructive interference of the two graphs in Fig. 2,
there are no terms in (\ref{M'}) proportional to
$1/M_W^2$.  This preserves unitarity.

A simple analytic expression for $\Gamma(W'\to \bar\ell\ell W)$ can be
obtained in the large $M_{W'}$ limit.  The result is
\begin{eqnarray}
\Gamma(W'\to \bar\ell\ell W) = {2g^2\Gamma(W'\to \bar ff')
\over 192\pi^2(1 + 3 \cot^4\phi)}
\Bigl[\Bigl(\ln{M_{W'}^2\over M_W^2}\Bigr)^2
- 5 \ln{M_{W'}^2\over M_W^2} - {\pi^2\over 3} +{37\over 3}
+{\cal{O}}\Bigl({M_W^2\over M_{W'}^2}\Bigr)\Bigr].\label{GammaW'}
\end{eqnarray}
The double log term in (\ref{GammaW'}) arises from the
interference of the two graphs in Fig. 2 in the kinematic region in which
$W$ is soft.  For $M_{W'}\sim 1$ TeV  numerical evaluation of
$\Gamma(W'\to \bar\ell\ell W)$ using  (\ref{rate}) and the analytic
formula (\ref{GammaW'}) are in excellent agreement with
less than a few percent difference.

Although the rare decay $W'\to \bar\ell\ell W$ is suppressed by
a factor of $\alpha/2\pi$ compared to $W'\to \bar\ell\nu_{\ell}$,
the double log factor provides an enhancement.
The observation of this logarithmic enhancement
has led to a series of diagnostic studies \cite{rfxiv,PACOF,HR}
on the properties of $Z'$.  The origin of these log factors
is related to the infrared and collinear singularities
of $S$-matrix elements, and is well known in QED and QCD.

To compare $W'\to \bar\ell\ell W$ with the basic process
$W'\to \bar\ell\nu_{\ell}$ we define a ratio
\begin{eqnarray}
R_{\rm lep}= {B(W'\to \bar\ell\ell W)\over
              B(W'\to \bar\ell\nu_{\ell})}.\label{ratio}
\end{eqnarray}
We plot the distribution of $R_{\rm lept}$
with respect to the invariant mass of the charged lepton pair for
$M_{W'}=1$ TeV  in Fig. 3.  In accordance with
the aforementioned logarithmic enhancement, the distribution
is clearly dominated by configurations in which the dilepton
invariant mass is large, implying that the $W$ is
soft and/or collinear.

The number of events for the process
$pp\to W^{\prime\pm}\to \bar\ell\ell W^{\pm}$ in the
narrow width approximation of $W'$ is given by
${\cal{L}}\sigma(pp\to W^{\prime\pm})B(W^{\prime\pm}\to\bar\ell\ell W^{\pm})$.
The results along with their typical statistical errors are summarized
in Table 2.  Again, they should be contrasted with number zero for
a right-handed $W'$.  Due to the large $W'$ gauge coupling
the numbers for the $W'$ rare decay are about one
order of magnitude larger than those of the corresponding
$Z'$ decays \cite{rfxiv,PACOF,HR}.

The signal of the
production of $W'$ followed by the rare  decay $W'\to \bar\ell\ell W$
is very clean.  The major background comes from the process
$pp\to W'\to WZ$, with $Z$ decaying into a charged lepton pair.
Although in the present model
the coupling for the interaction $W'\to WZ$  is suppressed
by $M_W^2/M_{W'}^2$, this background can be significant
because of the enhancement of $W'$ decaying into longitudinally
polarized $W$ and $Z$.
However, the background events can cleanly be eliminated by requiring
the invariant mass of the charged lepton system to be bigger than $100$ GeV.
This cut has been built into the numerical calculation.  The loss of
the signal associated with this cut is insignificant (a few percent),
as expected from the kinematic distribution of $R_{\rm lep}$ (Fig. 3).
Another source of background comes from the standard model
process $pp\to WZ\to W\ell^+\ell^-$.  However, requiring the
$WZ$ invariant mass  to be equal to $M_{W'}\pm 10$ GeV
already puts the total cross section $\sigma(pp\to WZ)$ below
$\sigma(pp\to W')$ for $M_{W'}\sim (1- 3 )$ TeV.
The background from the $WZ$
production  is thus eliminated by
employing the dilepton invariant mass cut.

In this paper we have shown that the processes
$pp\to W'W$ and $W'\to \bar\ell\ell W$ can be useful tests of
whether a $W'$ has a $V-A$ or $V+A$ coupling at future
hadron colliders.
To illustrate the idea   we have  considered a specific
example in which the $W'$ couples to $V-A$ currents.
For $M_{W'}\sim (1 - 3)$ TeV,
it is shown that the LHC and SSC can produce sufficient
numbers of events from these processes for such
a left-handed $W'$.  On the other hand,
the absence of such events
would be a clean signal that the $W'$ is right-handed.
In addition, the rates for the above processes allow for
a determination of the relative strength of the
$V-A$ gauge coupling of $W'$ to quarks and lepton.

\paragraph*{Acknowledgments~:}
This work was supported in part by the U.S. Department of Energy
 Grand No. DE-AC02-76-ERO-3071,
and an SSC Fellowship (J.L.) from Texas National Research
Laboratory Commission.

\newpage

\newpage
\centerline{\bf Figure Captions}
            \begin{figure}[h]
\vspace{5mm}
{\bf Figure 1.}
Feynman diagrams for the associated production process
$pp\to W'W$.
\label{f:production}

\vspace{5mm}
{\bf Figure 2.}
Feynman diagrams for the rare decay process $W'\to \bar\ell\ell W$.
\label{f:raredecay}

\vspace{5mm}
{\bf Figure 3.}
Distribution of $R_{\rm lep}$ defined in (\ref{ratio}) with respect
to the invariant mass of the charged lepton pair.
\label{f:distribution}

           \end{figure}
\newpage
\begin{tabular}{|l|l|llr|} \hline\hline
Collider& $W'$ mass &\multicolumn{3}{c|}{$pp\rightarrow W' W $}\\ \cline{3-5}
         & $(TeV)$  & $\phi=\pi/8$ & $\phi=\pi/4$ & $\phi=3\pi/8$ \\ \hline
        & $1$ & $1514\pm 39$& $6681\pm 82$ & $4213\pm 65$  \\
SSC     & $2$ & $131\pm 11$ & $579\pm 24$  & $365 \pm 19$  \\
        & $3$ & $29\pm 5$   & $128\pm 11$  & $81\pm 9$     \\ \hline
        & $1$ &$3200\pm 57$ &$14120\pm 119$& $8907\pm 94$  \\
LHC     & $2$ &$146\pm 12$  &$644\pm 25$   &$406\pm 20$    \\
        & $3$ &$15\pm 4$    &$67\pm 8 $    &$42\pm 6$      \\ \hline\hline
\end{tabular}

\bigskip\bigskip

{\bf Table 1.}  Number of events of
the process $pp\to W'W$,
with $W'$ subsequently decaying into leptons ($e$ and $\mu$),
 at the SSC and LHC.
The errors are  statistical.
\newpage

\begin{tabular}{|l|l|lll|llr|} \hline\hline
Collider&$W'$ mass &
  \multicolumn{3}{c|}{ $pp\to W^{\prime +}\to W^+\bar\ell\ell $}  &
  \multicolumn{3}{c|}{ $pp\to W^{\prime -}\to W^-\bar\ell\ell $} \\ \cline{3-8}
     &$(TeV)$ &  $\phi=\pi/8$ &   $\phi=\pi/4$ &
                                          $\phi=3\pi/8$ &
               $\phi=\pi/8$&$\phi=\pi/4$&$\phi=3\pi/8$\\ \hline
        & $1$ & $384\pm 20$   &  $1694\pm 41$  &   $1068\pm 33$     &
                         $219\pm 15$   &  $967\pm 31$   &   $611\pm 25$   \\
SSC     & $2$ & $ 79\pm 9 $   &  $348\pm 19$   &   $220\pm 15$      &
                $ 40\pm 6 $   &  $175\pm 13$   &   $111\pm 11$   \\
        & $3$ & $ 24\pm 5 $   &  $107\pm 10$   &   $ 68\pm 8 $      &
               $ 11\pm 3 $   &  $ 49\pm 7 $   &   $ 31\pm 6 $   \\ \hline
        & $1$ & $1065\pm 33$  &  $4703\pm 69$  &   $2965\pm 54$  &
                $ 513\pm 23$  &  $2263\pm 48$  &   $1427\pm 38$  \\
LHC     & $2$ & $ 128\pm 11$  &  $566\pm 24$   &   $ 357\pm 19$  &
                $ 51\pm 7$    &  $225\pm 15$   &   $ 142\pm 12$  \\
        & $3$ & $ 21\pm 5$    &  $ 95\pm 10$   &   $ 60\pm 8$    &
                $  7\pm 3$    &  $ 33\pm 6$    &   $ 21\pm 5$    \\
\hline\hline
\end{tabular}

\bigskip\bigskip

{\bf Table 2.}  Number of events of the
process $pp\to W^{\prime\pm}\to
\bar\ell\ell W^{\pm}$ at the SSC and LHC,  where $\ell=e$ and $\mu$.
The errors are statistical.

\end{document}